\documentclass[]{elsart}
\usepackage{amssymb}
\usepackage[dvips]{graphicx}

\begin{document}

\begin{frontmatter}

\title{The phonon dispersion of graphite revisited}

\author{Ludger Wirtz\corauthref{cor1}}
\author{and Angel Rubio}
\corauth[cor1]{Email: lwirtz@sc.ehu.es, Fax: +34-943-015600}

\address{Department of Material Physics, University of the Basque
Country, Centro Mixto CSIC-UPV, 
and Donostia International Physics Center (DIPC),
Po.~Manuel de Lardizabal 4, 20018 Donostia-San Sebasti\'an, Spain}

\begin{abstract}
We review calculations and measurements of the phonon-dispersion relation
of graphite. First-principles calculations
using density-functional theory are generally in good agreement with the 
experimental data since the long-range character of the dynamical
matrix is properly taken into account. Calculations with a plane-wave
basis demonstrate that for the in-plane optical modes,
the generalized-gradient approximation (GGA) yields 
frequencies lower by 2\% than the local-density approximation (LDA)
and is thus in better agreement with experiment.
The long-range character of the dynamical matrix limits the validity of 
force-constant approaches that take only interaction with few neighboring
atoms into account. However, by fitting the force-constants to the 
ab-initio dispersion relation, we show that the popular 
4th-nearest-neighbor force-constant approach yields an excellent fit
for the low frequency modes and a moderately good
fit (with a maximum deviation of 6\%) for the high-frequency modes.
If, in addition, the non-diagonal force-constant  for the second-nearest
neighbor interaction is taken into account, all the qualitative features
of the high-frequency dispersion can be reproduced and the maximum
deviation reduces to 4\%. We present the new parameters as a reliable 
basis for empirical model calculations of phonons in graphitic nanostructures,
in particular carbon nanotubes.
\end{abstract}


\end{frontmatter}

\section{Introduction}
The enormous amount of work on the vibrational spectroscopy of carbon 
nanotubes \cite{dressbook1,dressbook2} has also renewed the interest in the
vibrational properties of graphite.
Surprisingly, the debate about the exact phonon dispersion relation
and vibrational density of states (vDOS) of graphite is still not closed.
This was demonstrated by several recent publications: i.) Gr\"uneis et
al. \cite{gru02} reparameterized the popular 4th-nearest-neighbor force
constant (4NNFC) approach \cite{jish82,jish93,dressbook1} leading 
to pronounced changes in the dispersion relation.
ii.) Dubay and Kresse \cite{duba03} performed calculations using
density-functional theory (DFT) within the local-density approximation
(LDA) for the exchange-correlation functional. 
The calculations are in good agreement with earlier
DFT-LDA calculations \cite{kress95,miy95,pav96} and with phonon-measurements
by high-resolution electron-energy loss spectroscopy (HREELS)
\cite{oshi88,aiz90,sieb97} but deviate considerably from the 4NNFC approach.
iii.) Most recently, Maultzsch et al. \cite{mau04} have presented very 
accurate measurements
of the optical phonon modes along the directions $\Gamma - M$
and $\Gamma - K - M$ using inelastic x-ray scattering.
The measurements are accompanied by calculations using DFT in
the generalized gradient approximation (GGA) which yields slightly
softer optical phonon frequencies than the DFT-LDA calculations 
\cite{kress95,miy95,pav96,duba03,wir03,vit04,yana04} and improves marginally the
agreement with experiment. However, since the GGA calculations
are done with a basis-set consisting of localized orbitals
while the LDA calculations were performed using a plane-wave expansion,
it is not clear how much of the deviation stems from the difference
in basis set and how much stems from the different approximation
of the exchange-correlation functional.

The purpose of this paper is to review the available theoretical and 
experimental data. We present {\it ab-initio} calculations using the
LDA and GGA and show that the calculations are in very good agreement
with the vast majority of the experimental data-points.
We also provide a new fit of the parameters in the
widely used force-constant models. In many model-calculations, 
parameters are used that are based on a fit of only selected experimental
data. We perform, instead, a parameter fit to our {\it ab-initio} calculations.

The structure of the paper is as follows:
In section \ref{abicalc}, we describe the results of 
{\it ab-initio} calculations for the phonon-dispersion.
In order to assess the influence of the exchange-correlation
potential on the high-frequency modes, we perform calculations
using LDA and GGA both in the framework of a plane-wave
pseudopotential approach. We compare the results with previous plane-wave
calculations and calculations using localized orbitals.
In section \ref{sumexp}, we summarize the available experimental
data and make a comparison with the theoretical dispersion relations.
In section \ref{fca} we describe the empirical approaches
for the phonon calculations. The central quantity is the
dynamical matrix, which can be either fitted directly
through force-constants that describe the atom-atom interaction
up to nth-nearest-neighbor or which can be constructed
using the valence-force field (VFF) method of Aizawa et al. \cite{aiz90}.
We fit the parameters of the 4NNFC and VFF approaches to
the {\it ab-initio} dispersion relation. The parameters provide
a simple, yet quantitatively reliable, basis for phonon calculations
in carbon nanostructures, in particular  nanotubes (using the
proper curvature corrections for small diameter tubes \cite{dressbook1}).

\section{First-principles phonon calculations}
\label{abicalc}
The calculation of the vibrational modes by first-principles methods
starts with a determination of the equilibrium-geometry (i.e. the relative
atomic positions in the unit cell that yield zero forces
and the lattice constants that lead to a zero stress-tensor).
The phonon frequencies $\omega$ as a function of the phonon wave-vector 
${\bf q}$ are then the solution of the secular equation
\begin{equation}
\det \left|
\frac{1}{\sqrt{M_s M_t}} C_{st}^{\alpha\beta}({\bf q}) - \omega^2({\bf q}) 
\right| = 0.
\label{phonsec}
\end{equation}
$M_s$ and $M_t$ denote the atomic masses of atoms $s$ and $t$ and the
dynamical matrix is defined as
\begin{equation}
C_{st}^{\alpha\beta}({\bf q}) = \frac{\partial^2 E}{\partial u_s^{*\alpha}
({\bf q})
\partial u_t^{\beta}({\bf q})},
\label{dynmat}
\end{equation}
where $u_s^\alpha$ denotes the displacement of atom $s$ in
direction $\alpha$. The second derivative of the energy in Eq.\ (\ref{dynmat})
corresponds to the change of the force acting on atom $t$ in 
direction $\beta$ with respect to a displacement of atom $s$ in
direction $\alpha$:
\begin{equation}
C_{st}^{\alpha\beta}({\bf q}) = \frac{\partial}{\partial u_s^{*\alpha}({\bf q})}
F^{\beta}_t({\bf q}).
\label{dynmat2}
\end{equation}
Note the ${\bf q}$ dependence of the dynamical matrix and the atom
displacements. In an explicit calculation of the dynamical matrix
by displacing each of the atoms of the unit cell into all three 
directions, a periodic supercell has to be used which is commensurate with the
phonon wave length $2\pi/q$.
Fourier transform of the q-dependent dynamical matrix leads to the
real space force constant matrix $C_{st}^{\alpha\beta}({\bf R})$ where 
${\bf R}$ denotes a vector connecting different unit cells.

A phonon calculation starts with a determination of the
dynamical matrix in real space or reciprocal space.
In the force constant approaches, a reduced set of
$C^{\alpha \beta}_{st}({\bf R})$ are fitted in order to reproduce experimental
data (see section \ref{fca} below). 
The force constants can be calculated by
displacing atoms from the equilibrium position, calculating the total energy of
the new configuration and obtaining the second derivative of the energy
through a finite difference method. This is the approach
chosen in the {\it ab-initio} calculations of graphite phonons
in Refs.~\cite{kress95,miy95,san99,duba03,mau04,yana04}. In order to calculate
the dynamical matrix for different ${\bf q}$, a super-cell has to be chosen
that is commensurate with the resulting displacement pattern of the atoms.
An alternative is the use of density-functional perturbation theory DFPT 
\cite{baroni,gonze} 
where the atomic displacement is taken as a perturbation potential
and the resulting change in electron density (and energy) is calculated
self-consistently through a system of Kohn-Sham like equations. 
The main advantage is that one can compute phonons with
arbitrary ${\bf q}$, performing calculations using only a single unit-cell.
This method has been used in Refs.~\cite{pav96,wir03,vit04} and is used
for the calculations in this paper.
In both approaches, if the dynamical matrix is calculated on a sufficiently 
large set of ${\bf q}$-points, phonons for any ${\bf q}$ can be calculated 
by interpolating the dynamical matrix.
For many different materials (insulators, semiconductors, and metals)
phonon-dispersions with an accuracy of few cm$^{-1}$ have thus
been obtained \cite{baroni}.

The major breakthrough in the exact determination of the graphite-dispersion relation
were the first {\it ab-initio} calculations 
by Kresse et al. \cite{kress95} and Pavone et al. \cite{pav96}.
The calculations were done in the framework of DFT, employing the local-density 
approximation (LDA) to the exchange-correlation with a plane-wave expansion 
of the wavefunctions and using pseudo-potentials for the core-electrons.
These calculations introduced considerable qualitative changes in the behavior
of the high-frequency branches as compared to earlier force-constant fits.
In particular, these calculations established a crossing of the longitudinal and 
transverse optical branches along the $\Gamma-K$ as well as the $\Gamma-M$ direction
(see Fig.~\ref{abifig} below).
Since then, improvements in the computer codes, the use of better pseudo-potentials
and higher convergence-parameters have only led to small changes in the dispersion
relations obtained by codes using plane-wave expansion and pseudopotentials.
Slight variations mainly occur in the frequencies of the optical branches. 
Very recently, Maultzsch et al. presented
{\it ab-initio} calculations \cite{mau04} that are apparently in better agreement with 
experimental data. There are two sources of difference:
the use of the generalized-gradient approximation (GGA) to 
the exchange-correlation functional and the use of a localized-orbital basis set.

In order to demonstrate the high degree of convergence of the theoretical calculations
and in order to disentangle the influence of the exchange-correlation functional
from the influence of the basis-set on the high-frequency modes,
we have performed calculations both with the LDA and the GGA functional using a plane
wave expansion. The only parameter that controls the basis set is the energy cutoff.
Therefore, full convergence of the phonon frequency with respect to the basis set can be
easily tested by increasing the energy cutoff.

The calculations have been performed with the code {\tt ABINIT} \cite{abinit1,abinit2}.
We use a periodic supercell with a distance of 10 a.u. between neighboring
graphene-sheets. We checked that at this distance, the inter-layer interaction
has virtually no effect on the phonon frequencies. (Calculations on
bulk graphite are presented at the end of this section for completeness. 
Even there, the inter-plane interaction is so weak that it is only the
branches with frequencies lower than 400 cm$^{-1}$ that are visibly affected).
The dynamical matrix is calculated with DFPT \cite{baroni}. 
For the LDA functional we use
the Teter parameterization \cite{parlda} and for the GGA functional the
parameterization of Perdew, Burke, and Ernzerhof \cite{pargga}.
Core electrons are described by Troullier-Martins (TM) pseudopotentials 
\cite{trou}. For both LDA and GGA, an energy cutoff at 40 Ha. is used.
The first Brillouin zone is sampled with a 20x20x2 Monkhorst-Pack grid. 
We employ a 0.004 a.u. Fermi-Dirac smearing of the occupation around
the Fermi level. The phonon frequencies are converged to within 5 
cm$^{-1}$ with respect to variation of the energy cutoff and variation 
of k-point sampling. The influence of the smearing parameter is negligible.
The dynamical-matrix, which is the Fourier-transform of the real-space
force constants, is calculated on a two-dimensional 18x18 Monkhorst-Pack grid in
the reciprocal space of the phonon wave-vector ${\bf q}$. From this, the
dynamical matrix at any ${\bf q}$ is obtained by interpolation.
(We checked the quality of this interpolation by computing phonons for
some ${\bf q}$-vectors not contained in the grid and comparing
with the interpolated values).

The results of the calculation for the graphene sheet are presented 
in Fig.~\ref{abifig}. We compare with the LDA calculations of
Dubay and Kresse \cite{duba03} who used the Vienna {\it ab-initio} simulation
package ({\tt VASP}) \cite{vasp} with the projector 
augmented-wave (PAW) method \cite{paw} for the electron ion interaction.
Also shown are the GGA results of Maultzsch et al. who used the
{\tt SIESTA} package \cite{siesta} which also employs pseudopotentials for 
the core electron but uses a localized-orbital basis for the valence electrons.
In contrast to plane-waves there is no easy way to check the convergence
for a localized-orbital basis. Indeed, the converged value should
be the one obtained with the plane-wave basis set. Any difference can
be adscribed exclusively to the use of a localized basis set.

Before we analyze the differences between the different calculations,
we outline the features common in all {\it ab-initio} calculations of
graphite and graphene
\cite{kress95,pav96,san99,duba03,wir03,vit04,mau04,yana04}:
The phonon dispersion relation of the graphene sheet comprises three acoustic (A) branches
and three optical (O) branches. The modes affiliated with out-of-plane (Z) atomic motion
are considerably softer than the in-plane longitudinal (L) and transverse (T) modes.
While the TA and LA modes display the normal linear dispersion around
the $\Gamma$-point,
the ZA mode shows a $q^2$ energy dispersion which is
explained in Ref. \cite{dressbook1} as a consequence of the $D_{6h}$ point-group
symmetry of graphene.  Another consequence of the symmetry are the 
linear crossings of the ZA/ZO and
the LA/LO modes at the K-point. These correspond to conical intersections in the two-dimensional
parameter (${\bf q}$) space of the first Brillouin zone. Similarly, the 
electronic band structure of graphene displays a linear crossing at the 
K-point which marks the Fermi energy
and is responsible for the semi-metallic behavior of graphene.

For a meaningful comparison of phonon frequencies obtained by different 
calculations (using different pseudo-potentials, basis-sets,
parameterizations of the exchange-correlation functional), each calculation
should be performed at the respective optimized lattice constant.
For the discussion of the detailed differences between the calculations,
we present in table \ref{freqtable} the frequencies at the high-symmetry
points along with the respective optimized lattice constants. 
First, we compare our LDA calculation with the LDA calculation of
Dubay and Kresse \cite{duba03}.
While we obtain a lattice constant of 2.449 {\AA}, they obtain
slightly different values (depending on whether they use a soft
or a hard PAW). Nevertheless, the results for the
phonon frequencies at $\Gamma$ and M are almost identical with their
hard PAW results and only display minor differences ($\leq 1\%$)
from their soft PAW results. Apparently, small errors in the pseudopotential
that lead to small changes in the lattice constant are canceled in
the phonon-calculation which samples the parabolic slope of the
energy-hypersurface around the equilibrium position (at equilibrium
lattice constant). This hypothesis was confirmed by test-calculations
with other pseudo-potentials; e.g., a calculation with a 
Goedeker-Teter-Hutter potential \cite{gth} at an energy cutoff of
100 Ha., yielded a lattice constant of 2.442 a.u. and $\Gamma$-point
frequencies of 903 and 1593 cm$^{-1}$ (compared with the 893
and 1597 cm$^{-1}$ of the TM pseudopotential). In contrast, if we perform
a calculation with the Troullier-Martins pseudopotential at a lattice
constant which is slightly (0.4\%) enhanced with respect to the
optimized value, we obtain the frequencies listed
in table \ref{comptable}. These frequencies deviate by up to 2\% from the 
calculation at the optimized lattice constant.
We conclude that DFT-LDA calculations using plane-waves 
and performed at the {\em respective optimized lattice constant} can be 
considered well converged. Some differences remain only for the TO
mode around the K-point which seems to be most susceptible to 
variations of the pseudopotential/PAW parameterizations.
Besides that, all recent LDA calculations agree very well
with each other.

We quote four significant figures for the calculated lattice constant
because that is the order of convergence that can be achieved within the
calculations. Changes in the last digit lead to noticeable (1\%) changes 
in the phonon frequencies. 
However, it should be noted that the overall accuracy in comparison with
experimental lattice constants is much lower for two reasons:
(i) DFT in the LDA or GGA is an approximation to the exact $n$-electron
problem.
(ii) Temperature effects are neglected in the calculations, i.e., the 
calculations are performed for a fictitious classical system at zero
temperature.

The widely accepted value for the lattice constant of graphite at room 
temperature is $a_{RT} = 2.462$ {\AA} \cite{kelly,dress88}. Scaling to zero 
temperature according to the thermal expansion data of Bailey and Yates
\cite{bail70} yields $a'_{0K} = 2.455 ${\AA}. However, comparing this value
to the {\it ab-initio} value is, strictly speaking, not correct because
the {\it ab-initio} value neglects the anharmonic effect of the zero-point
vibrations. Instead, the {\it ab-initio} value should better be compared 
to the ``unrenormalized'' lattice constant at zero temperature, i.e.
to the value obtained by linearly extrapolating the temperature dependence
of the lattice constant at high temperature to zero temperature
(see Fig. 3 of Ref.~\cite{car01}). This ``unrenormalized'' value
corresponds to atoms in fixed positions, not subject to either
thermal or zero-point vibrations.
With the linear expansion coefficient 
$\left. \alpha \right|_{T=270K}=\left. \frac{1}{a}\frac{da}{dT}
\right|_{T=270K} = 1.27\time10^{-5}K^{-1}$ of Ref.~\cite{bail70},
we obtain $a_{0K}=2.452 ${\AA}. This value is between the LDA value 
$a=2.449${\AA}
and the GGA value $a=2.457${\AA} in agreement with the general trend
that LDA underestimates and GGA overestimates the bond-length.
We note that another value for the lattice constant that is sometimes
quoted in the literature is the value of Baskin and Meyer \cite{bask55}:
$a_{RT}=2.4589\pm0.0005 ${\AA} with a change less than 0.0005 {\AA}
as the specimen is cooled down to 78 K.

We turn now to the comparison of our LDA and GGA calculations.
Following our statement above, we present calculations at the
respective optimized lattice constants.
Fig.~\ref{abifig} and Tab.~\ref{freqtable} demonstrate very good agreement
for the acoustic and the ZO modes. The deviation hardly ever reaches
1\% of the phonon frequency. For the LO and TO modes,
the GGA frequencies are softer by about 2\% than the LDA values.
Particularly sensitive is the K-point where the softening of GGA
versus LDA reaches almost 3\% (37 cm$^{-1}$). However, in order to put
this effect of $v_{xc}$ into the right perspective, we note that the 
choice of the pseudo-potential (soft PAW versus Troullier-Martins) 
within the LDA approximation has a similarly big effect on the
TO mode at K as shown above.
Contrary to what is stated in Ref.~\cite{mau04}, the deviations
at the K-point do not arise from a neglect of the long-range character
of the dynamical matrix which is properly taken into account in the
supercell-approach (see also Ref.~\cite{duba03} where the
real space force constants are explicitly listed up to 20th-nearest-neighbor
interaction).
Compared to the experimental phonon-frequencies at the $\Gamma$-point
which can be determined with high accuracy by Raman-spectroscopy
\cite{nem77,tou70,bri71,fri71}, the GGA yields a slight underestimation
of the LO/TO mode while the LDA yields a slight overestimation.
For the ZO mode, both LDA and GGA overestimate the experimental
value by 4\% and 3\%, respectively.

Fig.~\ref{abifig} also displays the recent GGA calculation by Maultzsch 
et al.~\cite{mau04}.
In general, the agreement is very good with two exceptions:
i.) In our calculation the TO mode is about 2\% softer. ii.)
The localized-basis calculation yields at $\Gamma$ a ZO frequency 
of 825 cm$^{-1}$ which is considerably smaller than the Raman
value of 868$^{-1}$. The differences are entirely due to the
choice of the basis-set.

So far, we have only dealt with the single graphene sheet.
In Fig.~\ref{graphitefig}, we present a calculation of the dispersion
relation of bulk graphite. The calculation is done with DFT-LDA
using a 16x16x6 Monkhorst-Pack sampling of the first Brillouin zone.
The unit-cell of graphite (ABA stacking) contains 4 atoms which give
rise to 12 different phonon branches. However, as stated above,
for frequencies higher than 400 cm$^{-1}$, the phonon branches
are almost doubly degenerate since the inter-sheet interaction is
very week. The degeneracy is lifted because in one case, equivalent
atoms on neighboring sheets are oscillating in phase, while in the other case
they are oscillating with a phase difference of $\pi$. This gives
rise to small frequency differences of, in general, less than 10 cm$^{-1}$.
E.g., the calculated frequency difference between the
IR active $E_{1u}$ mode and the Raman active $E_{2g}$ mode at $\Gamma$ 
is 5 cm$^{-1}$ in perfect agreement with experiment (see table~\ref{freqtable}).
Since the branches are almost degenerate, the comparison with experimental
data can be done for phonon calculations of the graphene sheet only.

Only the phonon branches below 400 cm$^{-1}$ 
deviate noticeably from the branches of the sheet. They split into 
acoustic branches that approach 0 frequency for ${\bf q} \rightarrow 0$
(corresponding to in-phase oscillation of equivalent atoms of neighboring
sheets) and ``optical'' modes that approach a finite value (corresponding
to a phase-difference of $\pi$ in the oscillation of neighboring sheets).
The optimized lattice constant is 2.449 {\AA} as 
for the graphene sheet. As an optimized
inter-layer distance we obtain $l = 3.30$ {\AA} which is only slightly 
lower than the experimental value of $l = 3.34$ {\AA} (measured at 
a temperature of 4.2 K \cite{bask55}). This agreement is somewhat
surprising. The inter-layer distance is so large that the
chemical binding between neighboring-sheets (due to overlap of
$\pi$-electron orbitals) is assumed to be weak. Van-der-Waals
forces are expected to play a prominent role (up to the point that
occasionally the term "Van-der-Waals-binding" is used for the 
force that holds the graphite-sheets together). The Van-der-Waals
interaction is, however, not properly described, neither in the LDA
nor the GGA. (With GGA, we obtain an optimized lattice constant of 2.456 {\AA}
and a considerably overestimated inter-layer distance $l=3.90$ {\AA}). 
The good agreement of experimental and LDA theoretical
inter-sheet distance may therefore seem fortuitous. 
However, the detailed comparison of the low-frequency inter-layer modes
with neutron-scattering data \cite{nic72} demonstrate that also
the total-energy curve around the equilibrium distance is 
reproduced with moderately good accuracy in the LDA. 
This was already seen in the calculation
of Pavone et al.~\cite{pav96} and may be interpreted as an indication
that at the inter-layer distance, the chemical binding still
dominates over the Van-der-Waals force and only at larger inter-sheet distance
the Van-der-Waals force will eventually be dominant.
A more accurate description of the
low-frequency modes will be an important test for the design
of new functionals.

\section{Experimental data}
\label{sumexp}
In this section we give a summary of the available experimental data
for the phonon-dispersion relation of graphite. The data-points obtained
by different experimental methods are collected in Fig.~\ref{datafig}
and compared to our LDA and GGA calculations presented in the previous
section. 

{\em Inelastic neutron-scattering} is frequently used to obtain 
detailed information about the phonon-dispersion relation of crystalline
samples. Since it is very difficult to obtain large high-quality samples
of graphite, the available data \cite{dol62,nic72} is limited to the 
low-frequency ZA and LA branches (and the corresponding low frequency optical
modes ZO', TO', and LO'). The significance of the agreement of theory and
experiment for these branches has been discussed in the previous section.

{\em High-resolution electron energy loss spectroscopy} (HREELS) on graphite
and thin graphite-films \cite{wil87,oshi88,aiz90,sieb97,yana04}
has probed the high-symmetry directions $\Gamma$-K and $\Gamma$-M.
The measurements (data-points marked by squares in Fig.~\ref{datafig})
are consistent with each other and are in good
agreement with the calculations, taking the scattering of the
data points as a measure of the error bar.
However, one apparent discrepancy persists for the TA mode (also
called shear mode) around the M-point \cite{duba03} where the EELS
data converges towards 800 cm$^{-1}$ whereas the theory predicts
626 cm$^{-1}$ using LDA or 634 cm$^{-1}$ using GGA.
(The difference between calculations is much smaller than the difference
between theory and experiment).
The HREELS selection rules actually state that this mode should
be unobservable along the $\Gamma$-M direction owing to the reflection
symmetry \cite{ibach,yana04}.
Indeed, this branch was only observed in one experiment \cite{oshi88}
on bulk graphite and was not observed for experiments on thin-films
\cite{yana04}. The appearance of this branch (and the discrepancy
with respect to theory) may therefore tentatively
be explained as a consequence of limited crystalline quality with
the possible admixture of micro-crystallites of different orientation.
Therefore, it should not be used to fit force-field parameters (see
section \ref{fca} below). Instead, we will fit the parameters to the
first-principles calculations where no crossing between the TA and ZA
mode is present in the $\Gamma-M$ direction.

{\em Raman-spectroscopy} measures the phonon frequencies through the shift
in the wave-length of inelastically scattered photons. In first-order
Raman-scattering (one-phonon emission or absorption), only phonon-frequencies 
at the $\Gamma$-point can be detected, since the photons carry only
vanishing momentum compared to the scale of phonon momenta.
The selection rules of Raman-scattering, evaluated for the D$_{6h}$
point-group of graphite pose a further restriction.
The observable high-frequency mode \cite{nem77,tou70,bri71,fri71}
is the $E_{2g}$ mode at 1587 cm$^{-1}$.

{\em Infrared absorption spectroscopy} yields a value of 1587-1590 cm$^{-1}$ 
for the E$_{1u}$ mode and 868-869 cm$^{-1}$ for the A$_{2u}$ mode at $\Gamma$
\cite{nem77,kuhl98}.

In addition to the peaks due to symmetry allowed scattering-processes,
Raman spectra frequently display additional features such
as the disorder-induced $D$-band around 1350 cm$^{-1}$ (for laser
excitation at about 2.41 eV). 
This feature is strongly dispersive with the laser energy and is
explained as a $k$-selective resonance process \cite{mat99,poc98,fer01}.
A very elegant model is the {\em double-resonant Raman} effect proposed by
Thomsen and Reich \cite{tho00}. One possible scenario is a vertical resonant
excitation of an electron with momentum ${\bf k}$ close to the K-point, 
followed by an inelastic transition to another excited state with momentum
${\bf k}+{\bf q}$ under emission or absorption of a phonon, elastic 
backscattering to the original ${\bf k}$ mediated by a defect, and
de-excitation to the ground state by emitting a photon of different energy. 
The model was extended by Saito et al.\cite{sai02,gru02} to all branches
of the phonon dispersion relation and was used to evaluate the data of earlier
Raman experiments \cite{poc98,pim00,tan01,tan98,kaw95,kaw99,alv00,tan99}.
The results are depicted in Fig.~\ref{datafig} by asterisks.
The values close to $\Gamma$ are in fairly good agreement
with the HREELS data and the {\it ab-initio} calculations. In particular,
the values of the LO-branch coincide very well. A strong deviation
of the double-resonant Raman data from the calculations can be observed
at the $K$-point, in particular for the TO-mode. In the calculations, this
mode is very sensitive to the convergence parameters and to the lattice
constant (see tables \ref{freqtable} and \ref{comptable}). It may therefore
not come as a surprise that the presence of defects which is an essential
ingredient of the double-resonance Raman effect also yields a 
particularly strong modification of phonon frequencies for this branch
at the $K$-point. The presence of defects may also explain the strong deviation
of the double-resonant Raman data from both EELS-data and calculations
along the ZO-branch.

Using inelastic x-ray scattering, Maultzsch et al.~\cite{mau04} have
measured with high accuracy the high-frequency phonon branches.
The LO and TO branches along $\Gamma$-M and $\Gamma$-K are in good
agreement with the different HREELS measurements and are in almost
perfect agreement with the GGA-calculation. The most important achievement
is, however, that they experimentally established the dispersion relation 
along the line M - K. Also along this line, the agreement with
{\it ab-initio} calculations is quite good even though not as good
as along the other direction. Our calculations confirm the statement
of Ref.~\cite{mau04} that GGA yields a slightly better agreement than
LDA. Contrary to their calculation, however, we do not have to scale our 
theoretical results down by 1\% in order to obtain the good agreement.
This is because, we are using a fully converged plane wave basis set
instead of a localized basis set.
The remaining differences between theory and experiment may be due
to small deviations from the high-symmetry lines in the experiment.
At the same time, we recall that the GGA is still a very drastic
approximation to the unknown ``exact'' exchange-correlation functional.
Some of the remaining discrepancies could possibly be corrected by using
``better'' (as yet unknown) exchange-correlation functionals.

So far, limited crystal quality has prevented the measurement
of the full dispersion relation by means of neutron scattering. However,
neutron scattering on a powdered graphite sample has yielded
the generalized vibrational density of states (GvDOS) \cite{rols} of graphite.
The term "generalized" means that each phonon mode is weighted by
the cross section for its excitation.
The data is shown in Figure \ref{dosfig} a). We compare with 
the {\it ab-initio} vDOS \cite{vit04} (calculated with LDA) and the vDOS 
obtained from the model parameterization of Aizawa et al.~\cite{aiz90} and from
the 4NNFC approach of Ref.~\cite{jish93} (see next section about the
details of the models). The most pronounced peaks arise from
the high symmetry points as denoted in the figure and are in very
good agreement with the {\it ab-initio} vDOS. The experimental
DOS seems to confirm that the TA mode around the $M$-point has a
frequency lower than 650 cm$^{-1}$ which is in close agreement
to the {\it ab-initio} vDOS where the maxima arising from the TA($M$) and 
the ZO($M$) modes form one peak.

It can also be seen that the K-point phonons only contribute weak
peaks to the {\it ab-initio} vDOS between 900 and 1300 cm$^{-1}$
which are not resolved in the experimental vDOS. In contrast, the 
two model-calculations display a strong peak around 1200-1250 cm$^{-1}$
which arises from an incorrectly described LA mode along the line
$K-M$.

Recently, the vibrational density of states of graphite has also been measured
by inelastic scanning tunneling spectroscopy (STS) \cite{vit04}. The effect of 
phonon-scattering yields clear peaks at the corresponding energies in 
the second derivative of the I-V curve (Fig.~\ref{dosfig} b). Not all features
of the vDOS are resolved and additional inelastic scattering effects like
plasmon excitations seem to occur. However, the agreement with the vibrational
density of states is quite striking indicating that most (if not all)
phonon modes can be excited in STS.

In conclusion, the different experimental methods are in good agreement
with each other and yield a fairly complete picture of the phonon dispersion
relation and the vibrational density of states. The agreement with the
theoretical curves, in particular with the GGA calculation is very good.
Some small differences remain: On the experimental side, limited crystal
quality and the difficulty to perfectly align the crystal samples yield
some scattering of the data points. More important however may be the
role of temperature. While most experiments are performed at room temperature,
the calculations are performed in the harmonic approximation for a
classical crystal at zero temperature. While {\it ab-initio} calculations
for the temperature dependence of phonons in graphite are still missing,
recent calculations \cite{laz03} for MgB$_2$ have demonstrated that the 
high symmetry phonons at room temperature are softer by about 1\% than the 
phonons at zero temperature. Assuming a similar temperature dependence 
for graphite, the effect of temperature is of similar magnitude as the
difference between LDA and GGA. The small differences between the different
calculations should therefore be taken with caution whenever the quality
of the approximations are assessed by comparison with experimental data.

\section{Force constant approaches}
\label{fca}
We have shown in the previous sections that the major goal
of an accurate calculation of graphite-phonons in agreement with
experiment has been achieved. Nevertheless, for investigation
of carbon nanostructures (in particular, nanotubes) it is often
desirable to have a force-constant parametrization for fast - yet
reliable - calculations.
We review in this section the two main approaches found in the literature
on graphite phonons: the valence-force-field (VFF) model and the
direct parametrization of the diagonal real-space force-constants 
up to 4th-nearest-neighbor (4NNFC approach). We also give a new
parametrization of both models fitted to our first-principles calculations.

The general form of the force-constant matrix for the interaction of
an atom with its nth-nearest neighbor in the graphene sheet is
\begin{equation}
C_{n}=\left( \begin{array}{ccc}
\phi^l_{n}   & \xi_n     & 0    \\
-\xi_n    &  \phi^{ti}_{n} & 0  \\
0      & 0     &  \phi^{to}_{n} \\
\end{array} \right).
\label{general}
\end{equation}
The coordinate system is chosen such that $x$ is the longitudinal coordinate
(along the line connecting the two atoms),
$y$ the transverse in-plane coordinate, and $z$ the coordinate perpendicular
to the plane. The block-diagonal structure of the matrix reflects the
fact that in-plane and out-of-plane vibrations are completely decoupled.
In addition, $\xi_1=\xi_3=0$, due to the hexagonal-structure
of graphene, i.e. displacing an atom towards its first/third nearest
neighbor will not induce a transverse force on this atom.
Up to 4th-nearest neighbor, there are thus 14 free parameters to determine.

The 4NNFC approach (see, e.g., Ref.~\cite{dressbook1}) makes 
the additional simplifying assumption that off-diagonal elements can be 
neglected, i.e. $\xi_2=\xi_4=0$.
The force constant matrix describing the interaction between an atom
and its nth-nearest neighbor has then the form
\begin{equation}
C_{n}=\left( \begin{array}{ccc}
\phi^l_{n}   & 0     & 0    \\
0     &  \phi^{ti}_{n} & 0  \\
0      & 0     &  \phi^{to}_{n} \\
\end{array} \right).
\end{equation}
This means that a longitudinal displacement of an atom could only induce
a force in longitudinal direction towards its nth neighbor, and a transverse
displacement could induce only a transverse force.
This assumption reduces the number of free parameters to 12.

The valence-force-field model determines the parameters of the matrix
in Eq.~\ref{general} through the introduction of ``spring constants'' that
determine the change in potential energy upon different deformations.
The spring constants reflect the fact that a sp$^2$ bonded system
tries to locally preserve its planar geometry and 120 degree bond angles.
Aizawa et al \cite{aiz90}, have introduced a set of 5 parameters,
$\alpha_1,\alpha_2,\gamma_1,\gamma_2,$ and $\delta$.
The parameters $\alpha_1$ and $\alpha_2$ are spring constants
corresponding to bond-stretching,
$\gamma_1$ is an in-plane and $\gamma_2$ an out-of-plane
bond-bending spring constant, describing how the force changes
as the in-plane and out-of-plane component of the bond-angle changes.
In addition, the constant $\delta$ describes the restoring force
upon twisting a bond. 

For a good introduction to the approach, we refer
the reader to the appendix of Ref.\cite{aiz90}. 
Here, we just show as one example in figure \ref{springfig} the effect of the
force-constant $\gamma_1$. The potential energy corresponding to the
in-plane angle bending is
\begin{equation}
\frac{\gamma_1}{2}\left[
\left[\frac{({\bf u}_0-{\bf u}_1)\times{\bf r}_{10}}{|{\bf r}_{10}|^2}\right]_z 
- 
\left[\frac{({\bf u}_2-{\bf u}_1)\times{\bf r}_{12}}{|{\bf r}_{12}|^2}\right]_z 
\right]^2, 
\label{potenergie}
\end{equation}
where ${\bf u}_i$ indicates the displacement vector of atom $i$,
${\bf r}_{ij}$ is the relative mean position of atom $i$ from atom $j$,
and the subscript $z$ means the component perpendicular to the surface.

Evaluating the forces that arise from the potential energy terms,
the force-constant matrices for up to 4th-neighbor interaction
take on the form
\begin{eqnarray}
C_{1} & = &\left( \begin{array}{ccc}
\alpha_1   & 0     & 0    \\
0     &  \frac{9}{2d^2}\gamma_1 & 0  \\
0      & 0     &  \frac{18}{d^2}\gamma_2 \\
\end{array} \right),  \\
C_{2} & = &\left( \begin{array}{ccc}
\alpha_2+\frac{3}{4d^2}\gamma_1 &  \frac{3\sqrt{3}}{4d^2}\gamma_1   & 0    \\
- \frac{3\sqrt{3}}{4d^2}\gamma_1   &  -\frac{9}{4d^2}\gamma_1 & 0  \\
0      & 0     &  -\frac{3}{a^2}\gamma_2+\frac{1}{d^2}\delta \\
\end{array} \right),  \\
C_{3} & = &\left( \begin{array}{ccc}
0   & 0  & 0    \\
0   & 0  & 0  \\
0   & 0  &\frac{2}{d^2}\delta \\
\end{array} \right), \nonumber \\
C_{4} & = &\left( \begin{array}{ccc}
0   & 0  & 0    \\
0   & 0  & 0  \\
0   & 0  &-\frac{1}{d^2}\delta \\
\end{array} \right). \\
\end{eqnarray}
The constant $d = d_{C-C}$ denotes the bond-length of graphite.
In contrast to the 4NNFC parametrization, the
diagonal in-plane terms in the 3rd and 4th nearest neighbor 
interaction are zero. On the other hand, the 2nd nearest-neighbor interaction
has a non-diagonal term. As illustrated in Fig.~\ref{springfig},
the force acting on atom 3 upon longitudinal displacement of
atom 1 (keeping atom 2 fixed) has a longitudinal and transverse component. 
This is a consequence of the angular spring constant $\gamma_1$ that 
tries to preserve the 120 degree bonding.
The appearence of the off-diagonal term in the VFF-model is the reason
why this model with only 5 parameters can yield a fit of similar
quality as the 4NNFC parametrization with 12 parameters (see Fig.~\ref{allfits}
below).  An early VFF-model \cite{yosh56,young63} in terms of 
only 3 parameters for the intra-sheet forces gave a good fit of the slope 
of the acoustic modes (which in turn determine the specific heat) but 
cannot properly describe the dispersion of the high-frequency modes.

An {\it ab-initio} calculation of the real-space
force-constant matrices has confirmed the appearence of pronounced
off-diagonal terms \cite{duba03}. 
The interpretation of force-constants in terms of the VFF model is very 
instructive but limited to near neighbor interactions.
The {\it ab-initio} calculations have, in contrast,
demonstrated the long-range character of the dynamical matrix \cite{duba03}.
Possible extensions of the VFF-model would have to take into account
the effects of the complex
electronic rearrangement upon atomic displacement.
E.g., the longitudinal force-constant $\phi^l_3$ for the third-nearest
neighbor interaction is zero in the VFF model but turns out to be 
negative in the {\it ab-initio} calculations \cite{duba03}.
This is illustrated in Fig.~\ref{thirdnei}.
As atom 0 is pushed to the right (while atoms 1,1',2,and 2' are kept
at fixed position), atom 3 experiences a force to the left.
A similar behavior can be observed for the benzene ring.
A possible interpretation is that the change of the angle
$\eta_1$ (coming along with a small admixture of sp$^1$ and sp$^3$
hybridizations to the sp$^2$ bonding of atom 1) induces a change 
in of the bond between atom 1 and 2. This change, in turn, imposes the
same hybridization admixture to atom 2 and thereby tries to
keep the angle $\eta_2 = \eta_1$. The (in-plane) third-nearest neighbor
interaction could thus be expressed in a potential-energy term
of the form $~\frac{1}{2}c(\eta_2-\eta_1)^2$. However, instead of
adding additional degrees of complexity to the VFF-model, it is
easier  to fit the force-constants up to nth-nearest-neighbor
directly.

We have fitted the five parameters of the VFF model and the
12 parameters of the 4NNFC model to our GGA-calculation (see Fig.~\ref{abifig}).
Furthermore, we have performed a fit with 13 parameters, where, in addition
to the 4NNFC parameters, we allow for a non-zero off-diagonal parameter $\xi_2$
for the 2nd-nearest neighbor interaction. In table~\ref{tabparams}, we 
list the obtained parameters and compare with the parameterizations 
available in the literature. The resulting dispersion relations are
displayed in Fig.~\ref{allfits}. In all cases, we compare with 
our GGA-calculation which represents very well the bulk of the data-points
as was demonstrated in Fig.~\ref{datafig}.
The fit parameters were obtained by minimizing 
\begin{equation}
\chi^2 = \sum_{n=1}^N \sum_{i=1}^6 \left(
\omega_{i,n}^{\mbox{model}} - \omega_{i,n}^{\mbox{GGA}} \right)^2
\end{equation}
for the 6 phonon branches on $N = 237$ $q$-points along the high-symmetry
directions of the first Brillouin zone. The resulting standard-deviation
$\sigma = \sqrt{\chi^2/N/6}$ may serve as a measure of the quality of the
fit and is also listed in table~\ref{tabparams}. 

The 4NNFC parameterizations (Fig.~\ref{allfits} a and b) and the
VFF parameterizations (Fig.~\ref{allfits} e) that are available in the
literature reproduce very well the slope of the acoustic branches.
However, large deviations occur for the high-frequency modes,
in particular at the edge of the first Brillouin zone. E.g.,
the fit in Fig.~\ref{allfits} b) fails completely to reproduce
the crossings of the LO and TO branches along the lines $\Gamma-M$
and $\Gamma-K$, and the fits in a) and e) reproduce it only along the
line $\Gamma$-K. Our fit of the 4NNFC model (Fig.~\ref{allfits} c)
yields a major improvement (with a mean deviation $\sigma = 15.4$ cm$^{-1}$), 
however still does not reproduce the LO-TO crossing along $\Gamma-M$.
This is only achieved, if we include the off-diagonal term $\xi_2$ in 
the model. This gives only a slight improvement in terms of the
standard deviation ($\sigma = 13.5$ cm$^{-1}$) but leads to a qualitatively
correct ordering of the high-frequency modes also along the line $M-K$.
Clearly, for a very high-accuracy fit, a fourth-nearest-neighbor approach
is not enough. In particular, the TO phonon at the K-point is very sensitive
to the parametrization and can only be accurately described if the 
long-range character of the dynamical matrix is properly taken into 
account \cite{mau04}. However, our fit has an average deviation 
of only 1\% from the GGA-curve and a maximum deviation (at $K$) of 4\%.
For many practical calculations this accuracy is more than sufficient
and we therefore expect, that our fit to the {\it ab-initio} calculations
may be of some help in the future. Even the 4NNFC fit without the diagonal
term should be sufficient for many applications, provided that the
details of the high-frequency phonon branches along $M-K$ are not important.

We have also fitted the VFF-model to the GGA calculation
(Fig.~\ref{allfits} f). Since this model contains only five parameters,
it cannot compete in accuracy with the 4NNFC. In particular, since we
fit for phonons of the whole Brillouin-zone, the slope of the acoustic
modes around $\Gamma$ deviates from the correct value.

Our comparison of force-constant models shows that the parameterizations
in the literature display some strong deviations from the presumably
correct dispersion relation. Nevertheless, in many applications of these
models, these deviations are not of major concern.
E.g. for the calculation of the sound-velocity and the elastic constants,
only the slope of the acoustic branches at $\Gamma$ needs to be
described properly. Another example is the description of Raman
spectroscopy in carbon nanotubes. The Raman active modes of the tube can be
mapped onto phonon modes of graphene with a momentum close to zero for
large diameter tubes. Therefore, for first order Raman scattering, only
the dispersion close to $\Gamma$ needs to be well reproduced by the 
model. This is indeed fulfilled by the available force-constant fits,
as Fig.~\ref{allfits} demonstrates. However, for applications where
the whole Brillouin-zone is sampled (e.g. for the interpretation
of double-resonant Raman spectra) the present fit provides a considerably more
accurate description.

\section{Conclusion}
In the present work we have reviewed the experimental and theoretical
studies of the phonon dispersion in graphene and graphite. We have provided
a detailed discussion of the different approximations used in the first
principles calculations. In particular we have shown the effect of
the exchange-correlation potential $v_{xc}$ on the phonon-dispersion relation
for a calculation with a fully converged plane-wave basis.
The GGA yields phonons in the high-frequency region that are softer
by about 2\% than phonons calculated in the LDA.
We have demonstrated that for a consistent comparison of different
calculations (with different $v_{xc}$ or different pseudopotentials)
it is mandatory to perform the calculation at the respective optimized
lattice-constant. Under these conditions, recent LDA-calculations
using plane waves give very similar results and can be considered
fully converged (with some minor residual differences due to the
employed pseudopotentials). In Fig.~\ref{datafig}, where we have 
collected the available experimental data-points, obtained by
different spectroscopy methods, we have shown that the {\it ab-initio}
calculations reproduce very well the vast majority of the experimental data. 
The GGA yields a slightly better agreement for the high-frequency
branches than the LDA.

Concerning force-constant models, we have fitted a fourth-nearest neighbor
model to our GGA calculation and obtain very good agreement
between the model and first-principles calculations. Minor
discrepancies for the LO and TO branches (in particular close to the K-point) 
are related to the lack of long-range interactions in the model. 
This parametrization, in particular if the off-diagonal term for 
second-nearest neighbor interaction is taken into account,
provides a coherent description of the first principles
calculations and does not suffer from uncertainties related to different
experimental techniques. We hope that the model will be of use for further
calculations of phonons in carbon nanotubes and other nanostructures.

\section{Acknowledgment}
We acknowledge helpful communication with experimentalists:
D. Farias, J. L. Sauvajol, J. Serrano, L. Vitali, and H. Yanagisawa
and theorists: G. Kresse and M. Verstraete.
The work was supported by the European Research and Training Network COMELCAN
(Contract No. HPRN-CT-2000-00128) and the NANOQUANTA network of excellence
(NOE 500198-2). The calculations were performed at the computing center of 
the DIPC and at the European Center for parallelization of Barcelona (CEPBA).

% now the references.

% Tables
\begin{table}
\begin{tabular}{|lr|c|c|c|c|c|c|} \hline
\multicolumn{2}{|l|}{}  &   Ref.~\cite{duba03}   &  Ref.~\cite{duba03}&
this work   &  this work   & Ref.~\cite{mau04} & \\ \hline
\multicolumn{2}{|l|}{$v_{xc}$}  &   LDA & LDA & LDA & GGA & GGA & Experiment\\
\multicolumn{2}{|l|}{pseudo-potential}  & soft PAW  &  hard PAW &  
TM  &  TM  & TM & \\
\multicolumn{2}{|l|}{basis-expansion} &  plane-wave & plane-wave & plane-wave & 
plane-wave & local orbitals &            \\
\hline
\multicolumn{2}{|l|}{opt. lattice constant}&  2.451 {\AA} & 2.447 {\AA} & 2.449 {\AA} & 2.457 {\AA} & ? & 
2.452$^e$ {\AA} \\
\hline
$\Gamma$&ZO &  890 &  896 &  893 &  884 &  825 &  868$^a$   \\
        &LO/TO& 1595 & 1597 & 1597 & 1569 & 1581 &  1582$^b$,1587$^c$ \\
\hline
M  & ZA  &  475 &  476 &  472 &  476 &      &            \\
   & TA  &  618 &  627 &  626 &  634 &      &            \\
   & ZO  &  636 &  641 &  637 &  640 &      &            \\
   & LA  & 1339 & 1347 & 1346 & 1338 & 1315 & 1290$^d$   \\
   & LO  & 1380 & 1373 & 1368 & 1346 & 1350 & 1323$^d$   \\
   & TO  & 1442 & 1434 & 1428 & 1396 & 1425 & 1390$^d$   \\
\hline
K&ZO/ZA  &  535 &      &  535 &  539 &      &            \\
 &TA     &  994 &      & 1002 & 1004 &      &            \\
 &LO/LA  & 1246 &      & 1238 & 1221 & 1220 & 1194$^d$   \\
 &TO     & 1371 &      & 1326 & 1289 & 1300 & 1265$^d$   \\
\hline
\end{tabular}
\caption{Comparison of most recent DFT-calculations
of phonon frequencies (in cm$^{-1}$) at 
high-symmetry points in graphene. For the $\Gamma$-point we also 
list the experimental values obtained by infrared absorption and 
Raman scattering
($^a$Ref. \cite{nem77}, $^b\omega(E_{2g})$, Refs. \cite{nem77,tou70,bri71},
$^c\omega(E_{1u})$, Refs. \cite{nem77} and \cite{fri71}).
For the $M$ and $K$-points we list inelastic x-ray data 
of $^d$Ref.~\cite{mau04}. $^e$The experimental lattice constant is the
``unrenormalized'' zero temperature value (see text).}
\label{freqtable}
\end{table}

\begin{table}
\begin{tabular}{|lr|c|c||lr|c|c||lr|c|c|} \hline
&& $a_{opt.}$ & $a_{exp.}$ &&&$a_{opt.}$ & $a_{exp.}$&&&$a_{opt.}$ & $a_{exp.}$\\
\hline
$\Gamma$&ZO &  893 &  894 &$M$ &LA   & 1346 & 1337 & $K$&LO/LA & 1238 & 1225 \\
  &LO/TO & 1597 & 1575 &       &LO   & 1368 & 1350 &   &TO   & 1326 & 1298 \\
  &      &      &      &       & TO  & 1428 & 1404 &   &     &      &      \\
\hline
\end{tabular}
\caption{Comparison of high-frequency modes calculated with LDA at the
optimized lattice constant ($a_{opt.}=2.449$ {\AA}, as in Table \ref{freqtable}) 
and at the lattice constant $a=2.458$ {\AA}.
The small (0.4\%) change in the lattice constant affects strongly the
high-frequency modes (up to 2\% shift).}
\label{comptable}
\end{table}

\begin{table}[h]
\begin{center}
\begin{tabular}{|l|c|c|c|c|} \hline
\multicolumn{5}{|c|}{Force Constant Fits} \\
\hline
&  4NNFC & 4NNFC & 4NNFC  &  4NNFC\\
&  diagonal      &  diagonal     &  diagonal      & + off-diag. coupling\\
&  Ref.~\cite{jish93} & Ref.~\cite{gru02} & 
fit to GGA & fit to GGA  \\
\hline
$\sigma$ (cm$^{-1}$)      &  51.5    & 69.5  & 15.4  & 13.5  \\
\hline
$\phi^l_{1}$ (10$^4$ dyn/cm) &  36.50 & 40.37 & 39.87 & 40.98 \\
$\phi^l_{1}$ (10$^4$ dyn/cm) &   8.80 &  2.76 &  7.29 &  7.42 \\
$\phi^l_{1}$ (10$^4$ dyn/cm) &   3.00 &  0.05 & -2.64 & -3.32 \\
$\phi^l_{1}$ (10$^4$ dyn/cm) &  -1.92 &  1.31 &  0.10 &  0.65 \\
\hline
$\phi^{ti}_{1}$ (10$^4$ dyn/cm) & 24.50 & 25.18 & 17.28 & 14.50 \\
$\phi^{ti}_{1}$ (10$^4$ dyn/cm) & -3.23 &  2.22 & -4.61 & -4.08 \\
$\phi^{ti}_{1}$ (10$^4$ dyn/cm) & -5.25 & -8.99 &  3.31 &  5.01 \\
$\phi^{ti}_{1}$ (10$^4$ dyn/cm) &  2.29 &  0.22 &  0.79 &  0.55 \\
\hline
$\phi^{to}_{1}$ (10$^4$ dyn/cm) &  9.82 &  9.40 &  9.89 &  9.89 \\
$\phi^{to}_{1}$ (10$^4$ dyn/cm) & -0.40 & -0.08 & -0.82 & -0.82 \\
$\phi^{to}_{1}$ (10$^4$ dyn/cm) &  0.15 & -0.06 &  0.58 &  0.58 \\
$\phi^{to}_{1}$ (10$^4$ dyn/cm) & -0.58 & -0.63 & -0.52 & -0.52 \\
\hline
$\xi_2$         (10$^4$ dyn/cm) &  0    &   0   &   0   &  -0.91 \\
\hline \hline
\multicolumn{5}{|c|}{Valence-Force-Field Fits} \\
\hline
       & Ref.~\cite{aiz90} & Ref.~\cite{sieb97} &  fit to GGA & \\
\hline
$\sigma$   (cm$^{-1}$)      &  47.3 & 55.0 & 33.6 & \\
\hline
$\alpha_1$ (10$^4$ dyn/cm) &  36.4 & 34.4 & 39.9 & \\
$\alpha_2$ (10$^4$ dyn/cm) &   6.2 &  6.2 &  5.7 & \\
$\gamma_1$ (10$^{-13}$ erg) &  83.0 & 93.0 & 60.8 & \\
$\gamma_2$ (10$^{-13}$ erg) &  33.8 & 30.8 & 32.8 & \\
$\delta$   (10$^{-13}$ erg) &  31.7 & 41.7 & 34.6 & \\
\hline \hline
\end{tabular}
\end{center}
\caption{Force constant and valence-force-field parameterizations for graphene. 
Comparison of literature values with our fit to the GGA calculation.
The standard deviation $\sigma$ is calculated for each
parameterization with respect to the GGA calculation. The corresponding
dispersion relations are shown in Fig.~\ref{allfits}.}
\label{tabparams}
\end{table}

% Figures
\begin{figure}[htpb]
 \centering
   \includegraphics[draft=false,keepaspectratio=true,clip,%
                   width=0.85\linewidth]%
                   {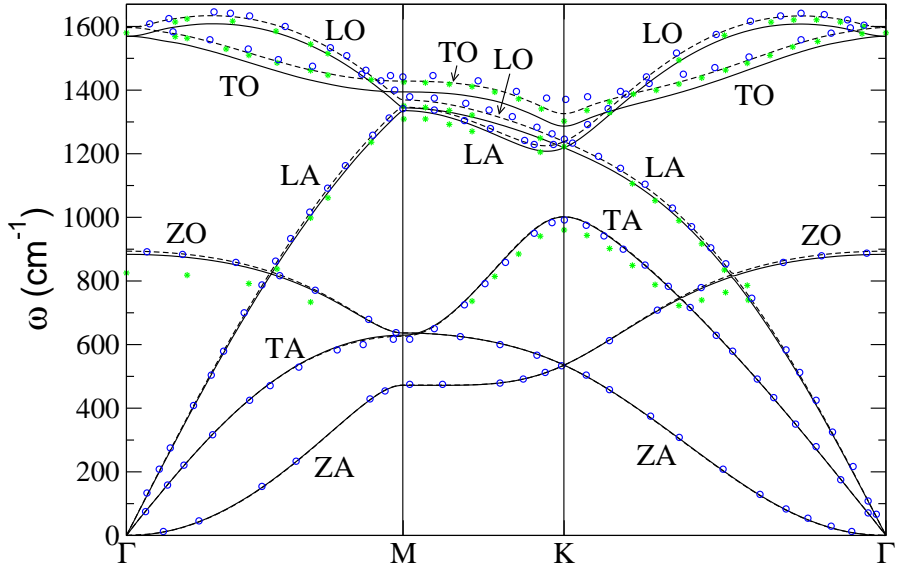}
\caption{{\it Ab initio} phonon dispersion relation of graphene.
Dashed line: LDA calculation, solid line: GGA calculation. We compare 
with LDA calculations by Dubay and Kresse \cite{duba03} (circles) and
GGA calculations by Maultzsch et al.~\cite{mau04} (asterisks).}
\label{abifig}
\end{figure}

\begin{figure}[htpb]
 \centering
   \includegraphics[draft=false,keepaspectratio=true,clip,%
                   width=0.85\linewidth]%
                   {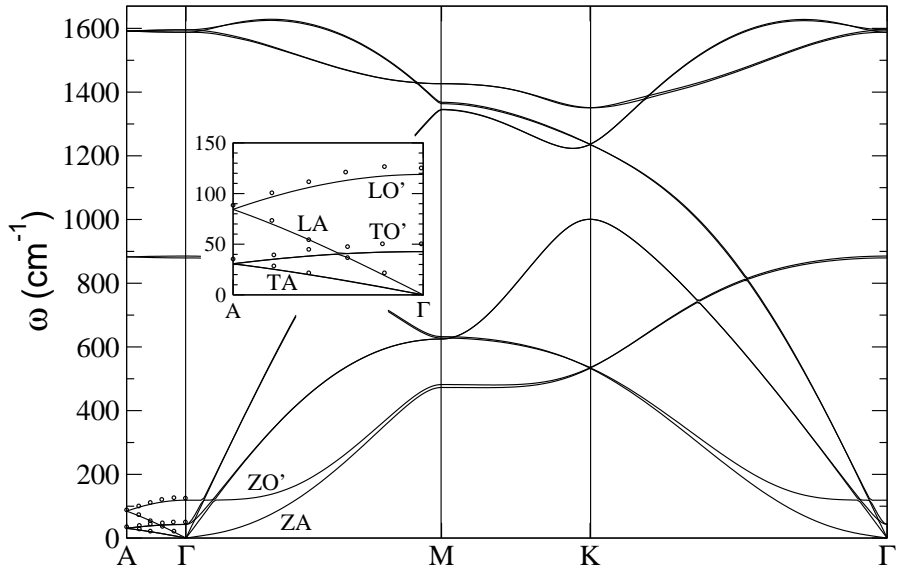}
\caption{{\it Ab initio} (LDA) phonon dispersion relation of bulk-graphite. 
In the inset, an enlargement of the low-frequency phonons along the
line $\Gamma$-A is shown.
Symbols denote the neutron-scattering data of Nicklow et al.~\cite{nic72}.}
\label{graphitefig}
\end{figure}

\begin{figure}[htpb]
 \centering
   \includegraphics[draft=false,keepaspectratio=true,clip,%
                   width=1.00\linewidth]%
                   {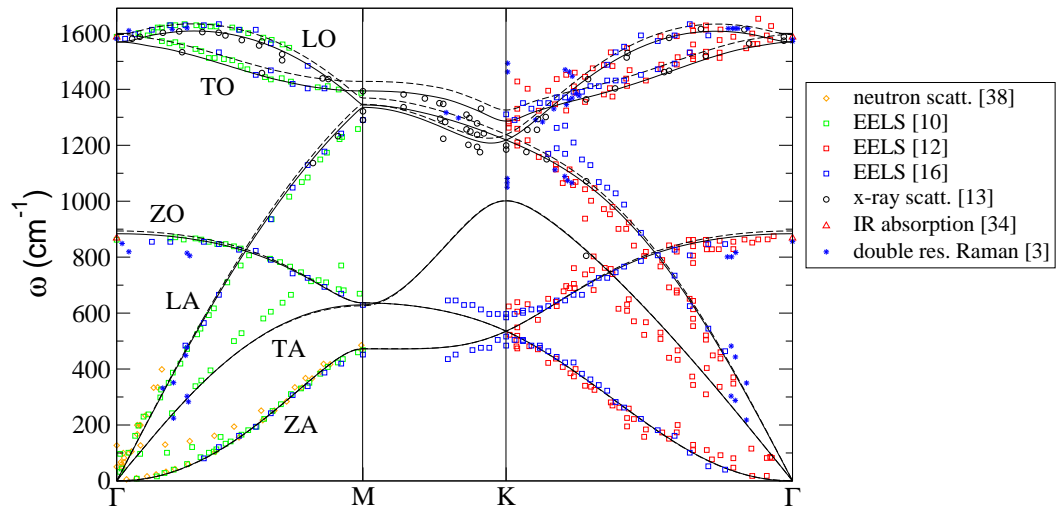}
\caption{(Color online): Experimental data points for the phonon-dispersion 
relation of graphite. Diamonds: neutron scattering \cite{nic72}, squares:
EELS \cite{oshi88,sieb97,yana04}, circles: x-ray scattering \cite{mau04},
triangles: IR absorption \cite{nem77}, asterisks: data of various
double resonant Raman scattering experiments, collected in Ref.~\cite{gru02}.
We compare with our ab-initio calculations: dashed line: LDA, solid line: GGA
(as in Fig.~\ref{abifig}).}
\label{datafig}
\end{figure}

\begin{figure}[htpb]
 \centering
   \includegraphics[draft=false,keepaspectratio=true,clip,%
                   width=0.85\linewidth]%
                   {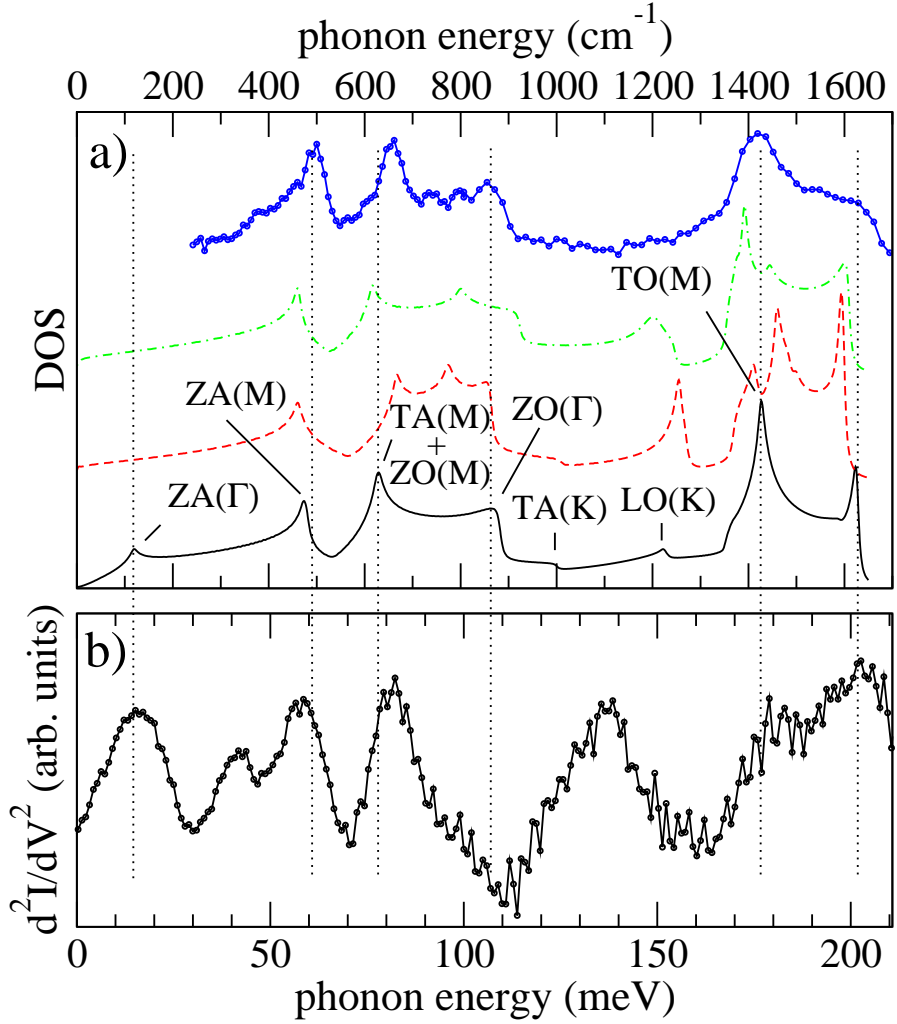}
\caption{a) Vibrational density of states (vDOS) of graphite.
Solid line: {\it ab-initio} calculation (LDA); 
dashed-line: 4NNFC model \cite{dressbook1};
dash-dotted line: model of Aizawa et al.~\cite{aiz90}; 
solid line with symbols: neutron-scattering on powdered sample \cite{rols}.
Note that the model calculations are for the graphene-sheet such that the
peak around 122 cm$^{-1}$ which is due to inter-plane coupling is missing.
b) second derivative of the I-V curve for inelastic scanning-tunneling
spectroscopy \cite{vit04} of a graphite-surface. For the assignments of
modes, see Fig.~\ref{abifig}.}
\label{dosfig}
\end{figure}

\begin{figure}[htpb]
 \centering
   \includegraphics[draft=false,keepaspectratio=true,clip,%
                   width=0.45\linewidth]%
                   {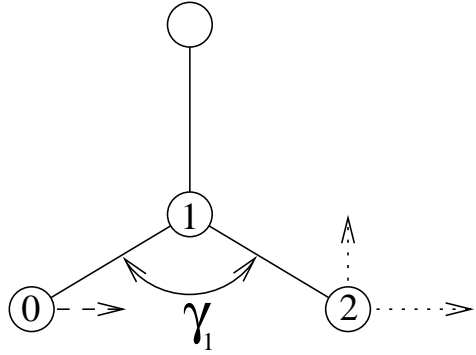}
\caption{Illustration of the in-plane bending spring constant $\gamma_1$.}
\label{springfig}
\end{figure}

\begin{figure}[htpb]
 \centering
   \includegraphics[draft=false,keepaspectratio=true,clip,%
                   width=0.45\linewidth]%
                   {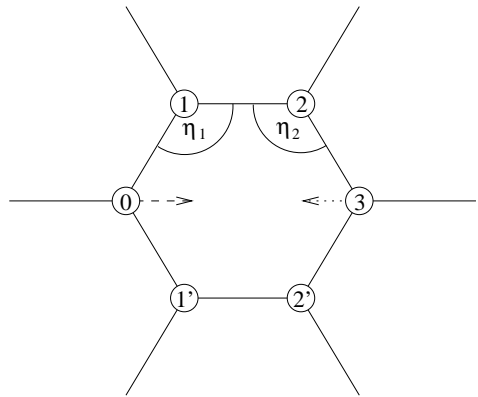}
\caption{Illustration of third-nearest neighbor interaction. Upon displacement
of atom 0 to the right and keeping all other atom fixed, atom 3 at the 
opposite corner of the hexagon experiences a force to the left.}
\label{thirdnei}
\end{figure}

\begin{figure}[htpb]
 \centering
   \includegraphics[draft=false,keepaspectratio=true,clip,%
                   width=1\linewidth]%
                   {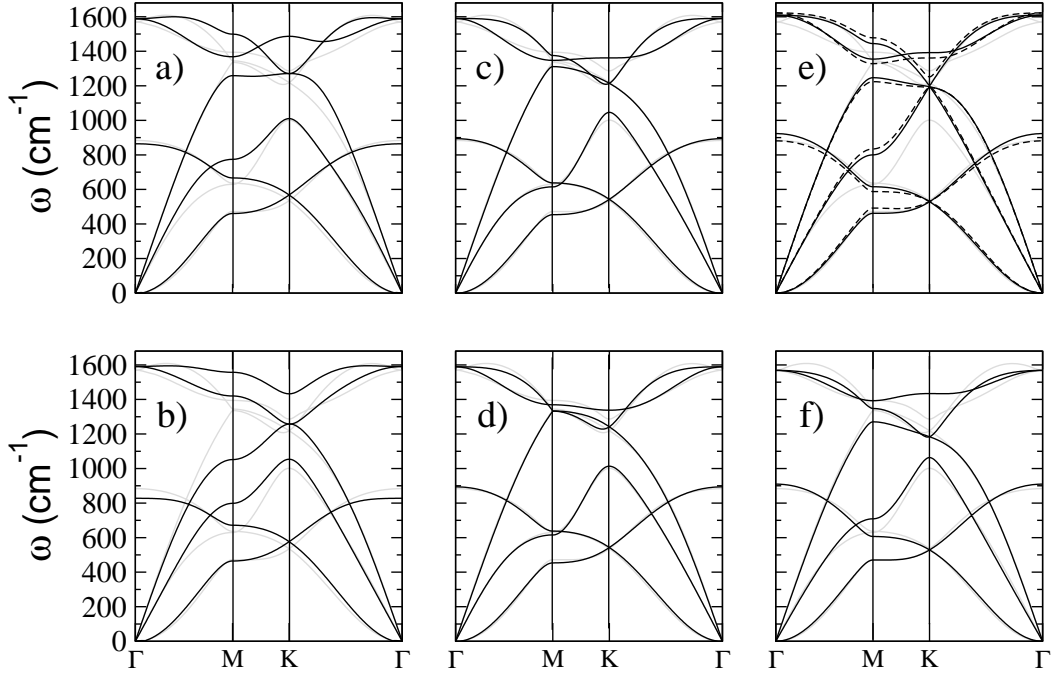}
\caption{Comparison of 4NNFC and VFF fits (black lines) with the GGA 
calculation (grey line). The corresponding parameters are 
listed in Tab.~\ref{tabparams}. 
a) 4NNFC fit of Ref.~\cite{jish93},
b) 4NNFC fit of Ref.~\cite{gru02} ,
c) Our 4NNFC fit to the GGA, 
d) Our 4NNFC fit to the GGA including off-diagonal term for the 2nd-nearest
neighbor interaction,
e) VFF fit of Ref.~\cite{aiz90} (solid line) and Ref.~\cite{sieb97}
(dashed line),
f) Our VFF fit to the GGA. Overall, the fits (c) and (d) clearly reproduce
most of the features of the phonon-dispersion relation of graphene. 
Notable exceptions are the TO mode at $K$ and the missing overbending of
the LO branch. These are related to the long-range nature of the
dynamical matrix and can only be properly reproduced with fits that take
into account the interaction of atoms more distant than fourth-nearest
neighbor.}
\label{allfits}
\end{figure}

\end{document}